\documentclass[10pt]{iopart}
\usepackage{epsfig,graphicx}

\newcommand{\nut}{\nu_t}

\begin{document}

\title{Spontaneous nucleation of structural defects in inhomogeneous ion chains}
\author{Gabriele De Chiara$^{1,2}$, Adolfo del Campo$^{3,4}$ , Giovanna Morigi$^{1,5}$, Martin B. Plenio$^{3,4}$, Alex Retzker$^{3,4}$}

\address{$^1$ Grup d'{\`O}ptica, Departament de F{\'i}sica, Universitat Aut{\`o}noma
de Barcelona, E-08193 Bellaterra, Spain}

\address{$^2$ F\'isica Te\`orica: Informaci\'o i Processos Qu\`antics, Universitat Aut\`{o}noma de Barcelona, E-08193 Bellaterra, Spain}

\address{$^3$Institut f{\"u}r Theoretische Physik, Albert-Einstein Allee 11, Universit{\"a}t Ulm, D-89069 Ulm, Germany}

\address{$^4$QOLS, The Blackett Laboratory, Imperial College London, Prince Consort Road, SW7 2BW London, UK}

\address{$^5$Theoretische Physik, Universit\"at des Saarlandes, D-66041 Saarbr\"ucken, Germany}

\def\d{{\rm d}}
\def\la{\langle}
\def\ra{\rangle}
\def\om{\omega}
\def\Om{\Omega}
\def\vep{\varepsilon}
\def\wh{\widehat}
\def\tr{\rm{Tr}}
\def\da{\dagger}
\newcommand{\beq}{\begin{equation}}
\newcommand{\eeq}{\end{equation}}
\newcommand{\beqa}{\begin{eqnarray}}
\newcommand{\eeqa}{\end{eqnarray}}
\newcommand{\intf}{\int_{-\infty}^\infty}
\newcommand{\into}{\int_0^\infty}

\begin{abstract}
Structural defects in ion crystals can be formed during a linear quench of the transverse trapping frequency across the mechanical instability from a linear chain to the zigzag structure. The density of defects after the sweep can be conveniently described by the Kibble-Zurek mechanism. In particular, the number of kinks in the zigzag ordering can be derived from a time-dependent Ginzburg-Landau equation for the order parameter, here the zigzag transverse size, under the assumption that the ions are continuously laser cooled. In a linear Paul trap the transition becomes inhomogeneous, being the charge density larger in the center and more rarefied at the edges. During the linear quench the mechanical instability is first crossed in the center of the chain, and a front, at which the mechanical instability is crossed during the quench, is identified which propagates along the chain from the center to the edges. If the velocity of this front is smaller than the sound velocity, the dynamics becomes adiabatic even in the thermodynamic limit and no defect is produced. Otherwise, the nucleation of kinks is reduced with respect to the case in which the charges are homogeneously distributed, leading to a new scaling of the density of kinks with the quenching rate. The analytical predictions are verified numerically by integrating the Langevin equations of  motion of the ions, in presence of a time-dependent transverse confinement.  We argue that the non-equilibrium  dynamics of an ion chain in a Paul trap constitutes an ideal scenario to test the inhomogeneous extension of the Kibble-Zurek mechanism, which lacks experimental evidence to date.
\end{abstract}
\maketitle

\section{Introduction}
%%%%%%%%%%%%%%%%%%%%%%%%%%%%%%%%%%%%
\begin{figure}
\begin{center}
\includegraphics[width=8cm,angle=0]{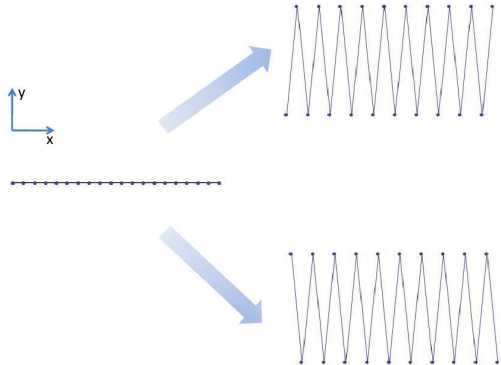}
\caption{
\label{transition}
Schematic representation of the ``linear to zig-zag''  phase transition in a homogeneous ion chain in two dimensions illustrating the structural phases involved. The high-symmetry phase corresponds to a linear chain, while the broken-symmetry phase is characterised by a doubly degenerate zig-zag configuration.
}
\end{center}
\end{figure}
%%%%%%%%%%%%%%%%%%%%%%%%%%%%%%%%%%%%

Ion crystals in Paul or Penning traps represent a prominent example of a self organized system that is amenable to accurate  experimental characterization and manipulation \cite{DubinRMP}. Ions confined  by means of static and radio-frequency electromagnetic potentials reach crystallization when laser cooled. Crystals made up from tens to millions of ions have been observed both in Paul traps \cite{MPQ,Raizen,Werth,Drewsen} and in Penning traps \cite{Bollinger}.
Different structures can be realized by varying the particle density or the trap anisotropy. In Ref.~\cite{MPQ} quasi one-dimensional structures have been experimentally characterized for first time. Here, the first three structures encountered when decreasing the transverse confinement are a linear chain, a planar zigzag structure, and a helicoidal arrangement of the ions about the trap axis. Phase transitions separating two structures are usually discontinuous. In this respect, the transition from the linear chain to the zigzag structure constitutes an exception. In fact, it has been demonstrated numerically \cite{Schiffer93,Piacente2005} and later analytically using Landau theory \cite{fishman2008} that the transition from a linear chain to the zigzag is of second order. The transition can be induced either by increasing the ion density or by decreasing the transverse confinement $\nut$ so that their value exceeds or is below, respectively, a certain critical value determining the mechanical instability. Recently, the transition linear-zigzag chain has been suggested as a test-bed for many-body quantum effects and the creation of double-well potentials \cite{RTSP}, and is well suited for the distinction between the nucleation of defects and quasiparticles in a symmetry breaking scenario \cite{Schuetzhold}.

\begin{figure}
\begin{center}
\includegraphics[width=12cm,angle=0]{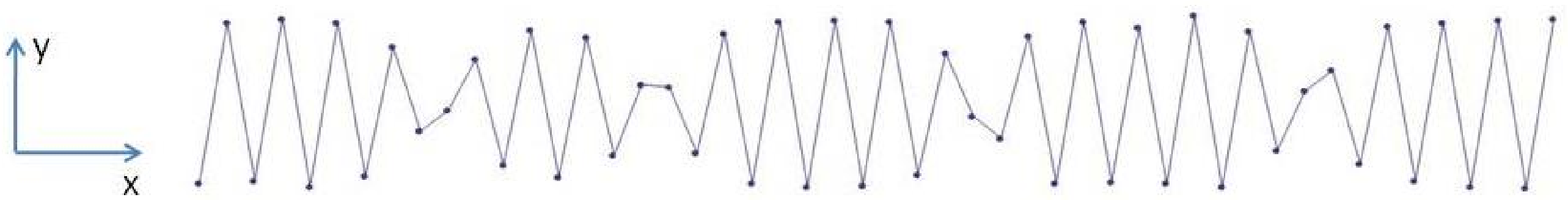}
\caption{\label{defectsring}
Ion chain in a zigzag configuration exhibiting structural defects. The ions are confined in a ring trap, and we report here the distribution of charges in two dimensions along the ring mapped to a line for clarity. The solid line joining the ions serves as a guide to the eye. In the example reported there are 4 defects. According to the Kibble-Zurek mechanism, the average size of the domains is given by the correlation length at the freeze-out time, $\hat{\xi}$ (see text for details).}
\end{center}
\end{figure}

The aim of this paper is to study the out-of-equilibrium dynamics of a linear chain of trapped ions when $\nut$ is lowered in time from above to below the mechanical instability separating the linear from the two-dimensional zigzag configuration, illustrated in Fig. \ref{transition}. Due to the finiteness of the sound velocity, space-like separated regions may develop one of the two possible zigzag orderings: odd ions up and even ions down or odd ions down and even ions up as depicted in Fig.~\ref{defectsring}. These regions are the analogs of magnetic domains in a ferromagnetic material and the interface between these domains is a structural defect.
The classical and quantum properties of these defects and their possible use for quantum information processing were studied in \cite{landa}. Defects formation can be understood from simple statistical mechanics considerations. If the rate of  change of $\nut$ is larger than the relaxation rate of the crystal, the latter does not have enough time to relax to the minimum energy configuration, which is characterized by a perfectly ordered zigzag structure, thus exhibiting no defects and corresponding to the state that would have been obtained after a perfect adiabatic transition. In principle, for an infinite system whose phase transition is described by Landau theory~\cite{Landau,HHreview}, the dynamical relaxation time scale diverges at the critical point and therefore no matter how slowly $\nut$ changes, there will always be proliferation of defects. This is the celebrated Kibble-Zurek mechanism (KZM) which can generally account for the nucleation of topological defects in scenarios of spontaneous symmetry breaking \cite{Kibble,Zurek}. Moreover, Zurek predicted the scaling of the number of defects as a function of the rate of passing through the phase transition \cite{Zurek}. The KZM scenario and Zurek's prediction for the scaling of the number of defects have been verified in a variety of systems numerically \cite{kzmnum} and experimentally \cite{kzmexp}. Recently, the KZ scaling prediction has also been extended to quantum phase transitions~\cite{kzmqpt}.

%%%%%%%%%%%%%%%%%%%%%%%%%%%%%%%%%%%%
%%%%%%%%%%%%%%%%%%%%%%%%%%%%%%%%%%%%

In this work we extend the results in our proposal \cite{ikzmions} and study the production of kinks in the linear to zigzag transition in ion crystals by deriving a time dependent Ginzburg-Landau theory for the transition. With this result, we determine the scaling of the number of defects with the rate of change of the transverse frequency which follows from the KZM and compare it with numerical simulations. In a ring trap when the interparticle distance is uniform, one recovers the standard KZM. By contrast, in a linear Paul trap the density of ions is inhomogeneous, higher at the center and more rarefied at the edges. When decreasing in time the transverse trap frequency, the value of the mechanical instability is first crossed in the center of the chain, and one can identify a front crossing the transition from the linear to the zigzag phase which moves from the center towards the edges, and whose velocity depends on the rate of the quench. If the velocity of this front is smaller than the sound velocity, at which a perturbation propagates along the chain, the dynamics becomes adiabatic. This result holds even in the thermodynamic limit.
Otherwise, the nucleation of defects is partially suppressed with respect to the homogeneous case and a novel scaling of the density of kinks with the rate of the transition emerges in the inhomogeneous extension of KZM \cite{Zurek09bec,ikzm}. The whole body of experimental results in \cite{kzmexp} has been aimed at the verification of the paradigmatic KZM in a homogeneous scenario (HKZM). By contrast, the inhomogeneous KZM (IKZM) \cite{Zurek09bec,ikzm} lacks to date experimental verification, and we argue in the following that ion chains in a Paul trap constitute an ideal system for such goal.

\section{Ginzburg-Landau equation in presence of laser cooling for the order parameter}

In this section we start from the theory presented in~\cite{morigi2004,fishman2008}, and derive a Ginzburg-Landau equation for the order parameter of the linear-zigzag structural transition. The order parameter corresponds to the position offset of the ions from the trap axis. For a large number of ions, using a local density approximation we can approximate the crystal as a continuum, so that the order parameter is a field. The theory is extended to the case in which the crystal motion is laser cooled, and the coupling to an external reservoir is described using the theoretical model developed in~\cite{morigi2001}. Defect formation by quenching the control parameter, here the transverse trap frequency, across the structural instability is studied by using the corresponding Langevin equation for the order parameter. This theoretical model allows us to estimate the density of defects at the end of the quench.

\subsection{Ginzburg-Landau Lagrangian for the structural phase transition}

The system we consider consists of $N$ ions of mass $m$, charge $Q$ and coordinates ${\bf r}_n=(x_n,y_n,z_n)$ which are confined in a quasi one dimensional trap with tight harmonic confinement at frequency $\nut$ in the plane $yz$. In this section we assume a ring trap with large radius $R$, so that we may impose periodic boundary conditions along the $x$-axis. The Lagrangian describing the dynamics of the ions is
\begin{equation}
\label{eq:discreteLangrangian}
L=T-V\,,
\end{equation}
where the kinetic and potential energies take the form, respectively,
\begin{eqnarray}
\label{Hamiltonian}
T&=&\frac{1}{2} m\sum_n \dot{{\bf r}}_n^2\,,
\\
V&=&\frac{1}{2} m\nu_t^2 \sum_n (y_n^2+  z_n^2) +\frac{Q^2}{2}\sum_{n\neq n'} \frac{1}{|{\bf r}_n- {\bf r_{n'}}|}\,.
\label{V}
\end{eqnarray}
At sufficiently low temperatures and sufficiently large transverse confinement the ions crystallize around the stable equilibrium points of the potential $V$, which are aligned along the $x$-axis, ${\bf r}_{0n}=(na,0,0)$, with $a$  the interparticle distance in the $x$-direction such that $a=2\pi R/N$.

The stability of the linear chain along the $x$ axis requires a transverse trap frequency exceeding a threshold value $\nu_t^{(c)}$, which scales with the characteristic frequency $\omega_0 =\sqrt{Q^2/ma^3}$. At $\nu_t^{(c)}$ the configuration has a structural instability, such that for $\nu_t<\nu_t^{(c)}$ the ions are organized in a planar structure with equilibrium positions ${\bf r}_{0n}=(na,(-1)^n (b/2)\cos\theta,(-1)^n (b/2)\sin\theta)$ with $\theta \in[0;2\pi]$ the angle between the crystal plane and the plane $xy$. The structure has the form of a zigzag line, which joins the charges along the plane, with $b$ the transverse size of the zigzag. An appropriate thermodynamic limit can be defined, letting the number of ions $N\to \infty$ while keeping the value $\nu_t^{(c)}$ and the characteristic frequency $\omega_0$ fixed. This corresponds to keeping fixed the linear interparticle distance $a$ or equivalently the charge density \cite{morigi2004}. One finds that the structural instability is a second-order phase transition, with control field $\nu_t$ (or alternatively, $a$) and order parameter $b$~\cite{fishman2008}. In particular, the critical value of the transverse frequency in the thermodynamic limit is given by $\nu_t^{(c)}=\omega_0\sqrt{7\zeta(3)/2}=(2.051\dots)\omega_0$, $\zeta$ being the Riemann-zeta function. From the Landau theory of the structural phase transition in~Ref.~\cite{fishman2008}, one can develop a continuum model for the field describing the order parameter applying standard assumptions, thereby obtaining the Ginzburg-Landau equation. We sketch below the assumptions made, in order to clarify the range of validity of the model which forms the basis of this work.

In the linear chain, in the harmonic limit, transverse and axial modes are decoupled. For periodic boundary conditions the modes have a well defined quasi-momentum $k$ which takes values within the Brillouin zone. For $N$ ions, the transverse, normal modes at wave vector $k$ are the real and imaginary part of the mode
\begin{equation}
\label{eq:normal}
\Psi_k^\sigma = \frac{1}{\sqrt N}\sum_n {\rm e}^{{\rm i}kna} \sigma_n
\end{equation}
where $\sigma_n=y_n,z_n$ is the transverse displacement of the ion at the equilibrium position ${\bf r}_{0n}$ and $k\in [0,\pi/a]$. 
Using decomposition \eref{eq:normal}, as shown in Ref.~\cite{fishman2008}, where all the details of the derivation are provided, the transverse potential, obtained by expanding Eq.~(\ref{V}) around the equilibrium positions, takes the form
\begin{equation}
\label{V:1}
V\simeq V^{(0)}+V^{(2)}+V^{(3)}+V^{(4)}
\end{equation}
where the label indicates the order of the expansion. In particular,
\begin{eqnarray}
\label{eq:V2}
V^{(2)}=\frac{1}{2}\sum_{k\in (0,\pi/a]}\sum_{\sigma=y,z}m\beta(k)\left({\rm Re}\{\Psi_k^{\sigma}\}^2+{\rm Im}\{\Psi_{k}^{\sigma}\}^2\right)
%+V^{(3)}+V^{(4)}
\end{eqnarray}
with
\begin{eqnarray}
\label{eq:beta}
\beta(k)=\nu_t^2-4\omega_0^2\sum_{j>0}\frac{1}{j^3}\sin^2\left(\frac{jka}{2}\right)
\end{eqnarray}
and where we omitted to write the potential term for the axial modes. The terms $V^{(3)}$ and $V^{(4)}$ correspond to third- and fourth-order expansion in the fluctuations around the equilibrium position, and contain the coupling between the axial and radial modes. We now make the assumption that the ions are pinned in the axial direction and can only oscillate in the transverse direction and discard the coupling to the axial modes. In this regime, it was found in Ref.~\cite{fishman2008} that  $V^{(3)} = 0$ and the fourth order term reads:
\begin{eqnarray}
V^{(4)}&=&\sum_{k_1+k_2+k_3+k_4=0} \sum_{\sigma,\tau=y,z}A(k_1,k_2,k_3,k_4)
\psi^\sigma_{k_1} \psi^\sigma_{k_2} \psi^\tau_{k_3} \psi^\tau_{k_4} 
\end{eqnarray} 
with
\begin{eqnarray}
\label {eq:Ak1}
A(k_1,k_2,k_3,k_4)=\frac{3}{2N}\frac{Q^2}{a^5}\sum_{m>0}\frac{1}{m^5}
\prod_{p=1}^4\sin{\frac{mk_pa}{2}}
\end{eqnarray}

In the stability regime one finds that $\beta(k)$ is minimum at $k_0=\pi/a$, corresponding to the so-called zigzag mode. The critical value of the transverse trap frequency $\nu_t^{(c)}$ is found from the relation $\beta(k_0)=0$. When $\nu_t< \nu_t^{(c)}$ the linear chain is unstable, and the ions order in a zigzag structure across the trap axis.

Hence, sufficiently close to the critical value $\nu_t^{(c)}$, the mode of the linear chain which first becomes unstable and determines the equilibrium positions of the zigzag structure (soft mode) is the transverse mode of the linear chain with the shortest wave length (zigzag mode), at wave vector $k_0=\pi/a$. When $\nu_t$ is sufficiently close to the critical value $\nu_t^{(c)}$, an effective potential can be derived for the transverse normal modes $\Psi_{\tilde k}^{\sigma}$ with wave vector $\tilde k=k_0-\delta k$, such that $a\delta k\ll 1$. The effective potential is composed  by a quadratic component, given in Eq.~(\ref{eq:V2}), with coefficient $\beta(\tilde k)$, Eq.~(\ref{eq:beta}) such that \cite{ramsey}
\begin{eqnarray}
\beta(\tilde k)\Bigl|_{\tilde k=k_0-\delta k} \approx \delta+h^2 \delta k^2,
\end{eqnarray}
where $\delta=\nu_t^2-\nu_t^{(c)2}$ and which can take negative values depending on the value of $\nu_t$. Within the same assumptions, we can approximate:
\begin{eqnarray}
A(k_1,k_2,k_3,k_4)\approx\frac{3}{2N}\frac{Q^2}{a^5}\sum_{m>0}\frac{1}{(2m-1)^5}
= \frac{m}{2N}\mathcal A,
\end{eqnarray}
where  ${\mathcal A}=(93\zeta (5)/32)\omega_0^2/a^2$ and
we obtain for the fourth order term:
\begin{eqnarray}
\label{eq:V4}
V^{(4)}&=& \frac{m}{2N}\mathcal A{\sum'_{\tilde k_1 + \tilde k_2+\tilde k_3+\tilde k_4=0}}\sum_{\sigma,\tau=y,z}   \psi^\sigma_{\tilde k_1} \psi^\sigma_{\tilde k_2} \psi^\tau_{\tilde k_3} \psi^\tau_{\tilde k_4}.
\end{eqnarray}
Note that the prime index in the sum over the quasimomenta refers to the fact that the sum is restricted to the quasi momenta $\tilde k_j=k_0\pm\delta k$, with $\delta k a\ll 1$. The potential form $V$ in Eq.~(\ref{V:1}) is valid at second order in the small parameter $\delta k a\ll 1$, when the amplitude of the transverse oscillations is much smaller than the interparticle distance $a$.

The limit $a\delta k\ll 1$ corresponds to a long-wavelength expansion made with respect to the zigzag mode: the phase difference of the considered transverse modes from the zigzag modes varies slowly from ion to ion, and a continuum description can be introduced. In this limit $x_n\to x$, where $x$ is a continuous variable, and $(-1)^n\sigma_n\to \psi^{\sigma}(x)$, where $\psi^{\sigma}(x)$ is now the position-dependent order parameter, which is related to $\psi_{\tilde k}^{\sigma}$ through the Fourier transform
\begin{equation}
\label{eq:fourier_cont}
\psi^\sigma_{\tilde k} = \frac{1}{\sqrt N} \int \frac{{\rm d}x}{a}\; e^{-{\rm i} \delta k x}   \psi^\sigma(x), \qquad \sigma=y,z
\end{equation}
while the factor $1/a$ in the right-hand side gives the density of states. For a large number of ions the discrete sum over $\tilde k$ vectors in the potential energy becomes an integral according to the rule $\sum_{\tilde k} \to \int {\rm d}(\delta k) Na/(2\pi)$.
The second order potential term becomes:
\begin{equation}
V^{(2)}=\frac{m}{2}\sum_\sigma \int \frac{d(\delta k) dx_1 dx_2}{2\pi a} (\delta+h^2\delta k^2)
e^{-i\delta k(x_1-x_2)} \psi^\sigma(x_1)\psi^\sigma(x_2),
\end{equation}
and using the integral representation of the Dirac function:
\begin{equation}
\int d(\delta k) e^{i\delta k x} = 2\pi\delta(x)
\end{equation}
we obtain:
\begin{equation}
V^{(2)}=\frac{m}{2}\sum_\sigma \int \frac{dx}{a} \left[\delta\psi^\sigma(x)^2 +h^2 \left(\partial_x\psi^{\sigma}(x)\right)^2\right].
\end{equation}
A similar calculations leads to:
\begin{equation}
V^{(4)}=\frac{m}{2}\mathcal A\sum_{\sigma,\tau} \int \frac{dx}{a} \psi^\sigma(x)^2 \psi^\tau(x)^2.
\end{equation}

Finally, we obtain the Lagrangian $L=\int {\rm d}x {\mathcal L}(x)$ where ${\mathcal L}(x)$ is the Lagrangian density and reads
\begin{eqnarray}
\label{Eq:GL}
{\mathcal L}(x)&=&\frac{1}{2}\frac{m}{a}\sum_{\sigma}\Bigl[\left(\partial_t\psi^{\sigma} (x)\right)^{2}-h^2\left(\partial_x\psi^{\sigma}(x)\right)^2
\nonumber\\
&-&\delta \psi^{\sigma}(x)^{2}-{\mathcal A}\psi^{\sigma}(x)^{2}\sum_{\tau}\psi^{\tau}(x)^{ 2}\Bigr].
\end{eqnarray}
We now discuss individually the parameters entering Eq. (\ref{Eq:GL}). The parameter
\begin{equation}
\delta=\nu_t^2-\nu_t^{(c)2},
\end{equation}
determines whether the ground state of the system is a linear chain or a zigzag, depending on its sign. The parameter $h=\omega_0a\sqrt{\log 2}$ is a velocity, and determines the speed with which a transverse perturbation propagates along the chain. Finally, the parameter $\mathcal A$ is positive and determines the value of the order parameter when $\delta<0$ as will be shown below. The Lagrangian density we derived has the form of a Ginzburg-Landau equation. It is valid for the modes of the linear chain close to instability and extends the theory presented in~\cite{fishman2008} for the specific case in which the coupling between axial and transverse modes can be discarded. The minimal energy solution of Eq.~(\ref{Eq:GL}) is constant in space and fulfills the relation:
\begin{equation}
\label{eq:psi_infinite}
\psi^{\sigma}[ \delta +2{\mathcal A} (\psi^{y 2}+\psi^{z 2})] = 0.
\end{equation}
Equation~(\ref{eq:psi_infinite}) always admits the solution $\psi^{\sigma} = 0$, corresponding to all the ions laying on the $x$ axis. This solution is clearly stable only for $ \delta>0$, i.e. in the linear chain phase. For $ \delta<0$ there is an infinite family of solutions satisfying $ \varrho = \pm \sqrt{- \delta/2{\mathcal A}}$, with $\varrho=\sqrt{\psi^{y 2}+\psi^{z 2}}$~\cite{fishman2008}. These solutions correspond to zigzag structures with the same amplitude but laid on different planes.

From the Ginzburg-Landau equation one can derive the critical exponent of the correlation length close to the critical point and at $T=0$. In particular, under the assumption of a small spatial and static deformation of the order parameter at a certain position, the field autocorrelation function at distance $x$ decays exponentially as $\sim \exp(-|x|/\xi)$~\cite{Landau}, where $\xi=h/\delta^{1/2}$ is the correlation length, such that
\begin{equation}
\label{xi}
\xi\sim a\left[\frac{\omega_0^2}{\nu_t^{(c)}\left(\nu_t-\nu_t^{(c)}\right)}\right]^{1/2}
\end{equation}
and it diverges as $\xi\sim (\nu_t-\nu_t^{(c)})^{-1/2}$ with the corresponding critical exponent is $1/2$. The same exponent governs the behaviour of the correlation length in the zigzag phase close to the critical point.

\subsection{Equation of motion of the slow modes in presence of laser cooling}

We now consider the physical situation, in which the crystal motion is laser cooled. For the moment, we consider the discrete distribution of ion charges forming a linear chain (so that the effective potential for the transverse modes is essentially quadratic). For Doppler cooling, an effective equation for the crystal modes can be derived, which has the form of a Fokker-Planck equation for the energy distribution of the normal modes~\cite{morigi2001}. This can be put in terms of a Langevin equation of the form
\begin{equation}
\label{eq:langevink}
\partial_t^2\psi_k^{\sigma}+\eta_{\sigma,k}\partial_t\psi_k^{\sigma}+\beta(k)\psi_k=\varepsilon_k^{\sigma}(t),
\end{equation}
which is here reported for the transverse modes. The damping rate $\eta_{\sigma,k}$ depends on the corresponding mode frequency and it can take different values depending on the propagation direction of the cooling lasers. The scalar $\varepsilon_k^{\sigma}(t)$ represents the corresponding Langevin force, such that its moments fulfill the relations
\begin{eqnarray}
\langle\varepsilon_k^{\sigma}(  t)\rangle&=&0,\\
\langle \varepsilon_k^{\sigma}(   t) \varepsilon_{k'}^{\sigma}( t')\rangle&=&2\eta_{\sigma, k}\kappa_BT \delta_{k,k'}\delta(  t-  t')/m,
\end{eqnarray}
where $\kappa_B$ is the Boltzmann constant and $T$ is the temperature which determines the thermal state due to cooling. Equation~(\ref{eq:langevink}) relies on the assumption that the dynamics of the electronic degrees of freedom is much faster than that of the motional degrees of freedom so they can be adiabatically eliminated from the equations for the degrees of freedom of the crystal normal modes, which is generally true when all ions are driven by the cooling laser~\cite{Footnote:2}.

In the following we focus on the situation in which the trap frequency along the $z$-axis is much larger than that along the $y$-axis. In this limit we drop the $\sigma$-label, and write an equation for the transverse modes along $y$ and close to the instability point, at the transition from a linear chain to a zigzag structure in the $xy$ plane. Here, the parameters $\eta_k$ and $\varepsilon_k$ are slowly varying and can be assumed to be constant. Under these assumptions, the Euler-Lagrange equation for $\psi=\psi^y(x)$ obtained from Eq.~(\ref{Eq:GL}) reads
\begin{equation}
\label{eq:TDGL}
\partial_t^2\psi-h^2 \partial_x^2\psi+\eta\partial_t\psi+\delta \psi+2{\mathcal A}\psi^3 = \varepsilon(t).
\end{equation}
 Note that $h$ can be obtained from the group velocity $s=\frac{h^2 k}{\sqrt{\delta+h^2k^2}}$ at the critical point $\delta=0$.
Damping introduces a characteristic relaxation time scale $\tau$, with which the system reaches equilibrium. Sufficiently close to the transition point, $\eta^2\gg \delta$ and
\begin{equation}
\label{tau}
\tau\approx \eta/\delta,
\end{equation}
see \cite{Landau} and the appendix.
Therefore the relaxation time $\tau$ diverges when the transverse frequency $\nu_t$ approaches the critical value $\nu_t^{(c)}$ as $\tau\sim (\nu_t-\nu_t^{(c)})^{-1}$.
The same scaling is valid also in the zigzag phase $\delta<0$. The quantities $\xi$ and $\tau$ and their scaling with $\delta$ are crucial for determining the scaling of the defect production during a quench of the transverse frequency as explained in the next section.

\subsection{Quenching the trap frequency through the critical value}

Within the Ginzburg-Landau description we now assume that the transverse trap frequency $\nu_t$ undergoes a change in time in the interval $[-\tau_Q,\tau_Q]$, such that 
\begin{equation}
\nu_t=\sqrt{\nu_t^{(c)2}+\delta(t)},
\end{equation}
and
\begin{equation}
\label{Eq:quench}
\delta(t)=-\delta_0\frac{t}{\tau_Q},
\end{equation}
with $\delta_0>0$ and $\delta_0\ll \nu_t^{(c)}$. The transverse trap frequency value is swept through the mechanical instability of the linear-zigzag chain, such that $\delta(-\tau_Q)=\delta_0>0$ and $\delta(\tau_Q)=-\delta_0<0$. Correspondingly, the equation for the field now reads
\begin{equation}
\partial_t^2\psi-h^2\partial_x^2\psi+\eta\partial_t\psi+\delta(t)\psi+2{\mathcal A}\psi^3=\varepsilon(t),
\end{equation}
where for a given time $t$ the ground state and thermodynamic properties at the phase transition are well defined. The time-dependence of $\delta(t)$ gives now a non-equilibrium problem. In principle, this should be solved considering the differential equation with time-dependent parameters. Nevertheless one might reach an understanding of the essential features by separating it into two domains. In the first domain, one can use the equilibrium solution provided that one may assume the adiabatic approximation, namely, $|\delta(t)/\dot{\delta}(t)|\gg \tau(t)$, where $\tau(t)$ is the relaxation time for the equilibrium situation at the given value $\delta=\delta(t)$.
The second domain is known as the {\it impulse region}, where as a result of the critical slowing down the order paramer is assumed to be frozen, ceasing to react to the external quench \cite{Zurek}.
This stage is separated from the adiabatic regime by the freeze-out time scale $\hat{t}$, i.e., a time scale in which the adiabatic condition ceases to be valid. This time scale can be estimated by setting $|\delta(\hat{t})/\dot{\delta}(\hat{t})|= \tau(\hat{t})$, which clearly sets a lower limit to $\hat{t}$. On this time-scale the order parameter is correlated over domains of characteristic length $\hat{\xi}=\xi(\delta(\hat{t}))$ given by the correlation length at the freeze-out time. The density of defects $d$ (number of defects over the total number of ions) can then be estimated by the relation $d\sim 1/\hat{\xi}$.

We now determine the density of defects for the chosen time variation of the parameter $\delta$, Eq.~(\ref{Eq:quench}). In this case, an estimate for the freeze-out time $\hat{t}$ is given by the instant of time $\hat t$ at which the rate of change of $\delta$ equals the relaxation time:
\begin{equation}
\label{hat:t}
%\hat{t}=\tau(\delta(\hat{t})).
\left|\frac{\delta(\hat{t})}{\dot{\delta}(\hat{t})}\right|=\hat t=\tau(\hat{t})
\end{equation}
Equation~(\ref{hat:t}) allows for simple solutions in two specific limits, which will be analyzed in this paper. 
In the limit, in which $\sqrt{\delta(\hat{t})}\ll \eta$ or equivalently $\eta^3\gg\delta_0/\tau_Q$, the damping overcomes the oscillations associated with the frequency at the freeze-out time scale, $\sqrt{\delta(\hat{t})}$. Note that this condition is imposed at the freeze-out-time before crossing the transition. 
It is then when the properties of the broken symmetry phase are determined. 
We denote this regime by ``overdamped limit'', following the definition introduced in Ref.~\cite{laguna1998}. 
In this case, using Eqs.~(\ref{Eq:quench}) and~(\ref{tau}) one finds $\hat{t}=\sqrt{\tau_0\tau_Q}$ with $\tau_0=\eta/\delta_0$, which sets $\delta(\hat{t})=\sqrt{\eta\delta_0/\tau_Q}$. The density of defects in the overdamped limit takes the form
\begin{equation}
\label{HKZMO}
d_{\rm o}\sim \frac{1}{\hat{\xi}_o}=\frac{1}{a}\frac{1}{\omega_0}\left(\frac{\delta_0\eta}{\tau_Q}\right)^{1/4}.
\end{equation}
where we used Eq.~(\ref{xi}), assuming that the temperature is sufficiently low to neglect finite temperature effects. Another limit, which we will consider, is the one in which $\sqrt{\delta(\hat{t})}\gg\eta$, namely, the Ginzburg-Landau equation is the one of an underdamped oscillator at the freeze-out time. We denote this regime by ``underdamped'' limit according to Ref.~\cite{laguna1998}. In this case, $\hat{t}=(\tau_0^2\tau_Q)^{1/3}$ with $\tau_0=1/\sqrt{\delta_0}$, $\delta(\hat{t})=(\delta_0/\tau_Q)^{2/3}$, and the density of defects is given by
\begin{equation}
\label{HKZMU}
d_{\rm u}\sim \frac{1}{\hat{\xi}_u}= \frac{1}{a}\frac{1}{\omega_0}\left(\frac{\delta_0}{\tau_Q}\right)^{1/3}.
\end{equation}
The algebraic scaling in Eq.~(\ref{HKZMU}) coincides with the one found in Ref.~\cite{laguna1998}. The different power scaling of $\delta_0$, with respect to Ref.~\cite{laguna1998}, is due to the fact that in the present expressions we have written explicitly the proportionality factors (which in~\cite{laguna1998} are summarized in the parameter $\xi_0$). We finally note that, the fact that the density of defects exhibits a power-law behaviour as a function of $\tau_Q$ is obviously due to the specific choice of the quench as a function of time and of the considered regime (determined by the ratio $\eta/\delta(\hat{t})$). The result can be understood as the domain size, determining the density of defects is given by the speed at which a perturbation propagates along the chain, $a \omega_0$, multiplied by the relaxation time scale.

The scenario just discussed describes the dynamics of the structural phase transition for ions placed on a ring with equilibrium positions ${\bf r}_{0n}$, where the interparticle spacing is homogeneous.
%We now verify numerically the scalings Eq.~(\ref{HKZMO}-\ref{HKZMU}) for the over and underdamped regimes
We now verify numerically the HKZM  prediction by integrating numerically the Euler-Lagrange equations obtained by minimizing the Lagrangian in Eq. (\ref{eq:discreteLangrangian}) for a finite number of ions, and whose motion is damped in presence of a Langevin force. Assuming a ring confinement and pinning of one ion, the relevant degrees of freedom in the critical region are the transverse coordinates $y_n$.
%%%%%%%%%%%%%%%%
\begin{figure}
\begin{center}
\includegraphics[width=8cm,angle=0]{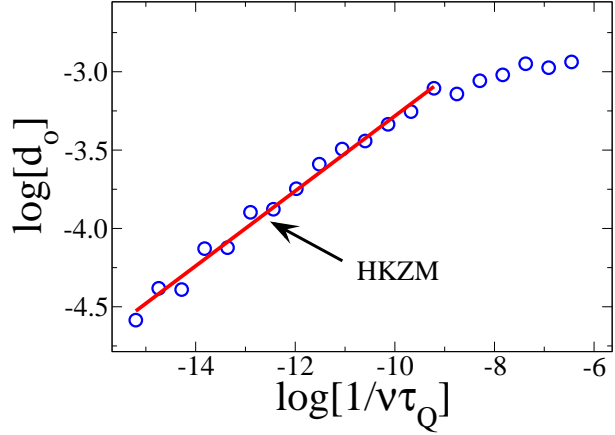}
\caption{\label{kzmring}
Scaling of the density of defects for in  an ion chain in a ring trap as a function of the rate of quenching the frequency of the transverse confinement.
The fit is $\log d= -0.889+0.239\log r$ (regression coefficient $0.994$) for $N=50$, and it involves average over $200$ realizations.
The parameters are $\eta=185\omega_0$, $\nu_t^{(c)}=2.05\omega_0$, $\delta_0=0.68\omega_0^2$, $\epsilon=2.5\times 10^{-3} a \omega_0^{3/2}$.
$l=Na$ is the length of the trap and $\epsilon$ is the parameter which controls the strength of the noise which is chosen by  adding the following term to the equation of motion $\epsilon N(0,1) \sqrt{\Delta t},$ where $\Delta t$ is the time step. This noise corresponds to a temperature of $k_B T=6.28 \times 10^{-6}\frac{ma^2\omega_0^{3}}{\eta}$, which in our case corresponds to a temperature of the order of $20$ phonons.  Note that the ratio of the number of defects over the total number of ions $N$ is related to the spatial density of defects by a factor $l/N$.
}
%
%$(\nu_t^{(c)})^2=1.232\nu^2$
\end{center}
\end{figure}
%%%%%%%%%%%%%%%%%%%%%%%%%%%%%%%%%%%
At $t=0$ the ions are in a linear chain configuration with $\langle y_n\rangle =0$.
The system is then driven by a linear quench of the form given in Eq.~(\ref{Eq:quench}), and the density of defects $d$  is computed at some asymptotic time, after which $d$ remains practically constant. A typical evolution instance with $4$ defects is shown in Fig.~\ref{defectsring}. These defects resemble the non-massive kinks of the Frenkel-Kontorova model with a transversal degree of freedom that can be described by an effective $\phi^4$ theory for the translational displacement \cite{FKM}.  The classical and quantum behavior of these kinks was studied in \cite{landa}.

In the numerics, for each set of parameters, the density of defects is averaged over many different realizations such as the one in Fig. \ref{defectsring}. We then study the dependence of $d$ on $\tau_Q$. A least-squares fit to the list of data is used to compute the scaling exponent. Figure \ref{kzmring} shows the scaling of the density of defects $d$ as a function of the frequency quench time $\tau_Q$ in the overdamped regime, in agreement with the homogeneous KZM. Deviations from KZM  occur for fast quenches, for which the density of defects saturates as a result of the interactions between different kinks. In addition, finite size effects  limits also the validity of the KZM scaling for very slow quenches that can lead to an adiabatic dynamics. Nucleation of defects is expected to be suppressed whenever the correlation length at the freeze-out time scale exceeds the size of the system, $\hat{\xi}>Na$, both in the underdamped and overdamped regimes.

\section{Ion chain inside a linear Paul trap: inhomogeneous effects}

\begin{figure}[t]
\begin{center}
\includegraphics[width=10cm,angle=0]{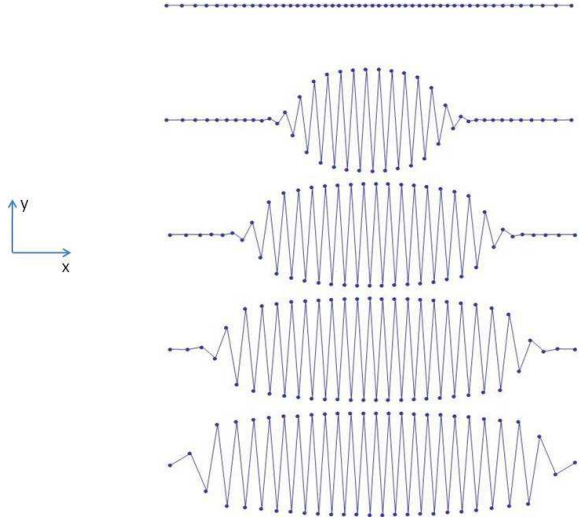}
\caption{\label{zztrap}
Sequence of classical ground states of a harmonically trapped 1D ion crystal for decreasing values of the transverse trap frequency $\nu_t$ (from top to bottom), across the mechanical instability from a linear to a zigzag chain. Due to the harmonic axial potential, the density of ions in the center is larger. Correspondingly, this is the first region where the zigzag structure is formed when decreasing $\nu_t$.
}
\end{center}
\end{figure}

The standard scaling of topological defects for homogeneous phase transitions should be revised whenever
the quench is local or there is a spatial dependence of the relative frequency $\delta$ \cite{ikzm,Zurek09bec}. In this section we discuss the dynamics of the structural phase transition for ion chains with open boundaries, for which both transverse and axial trapping potentials are harmonic. The ion chain in a linear Paul trap is characterized by a linear charge distribution which is inhomogeneous, with ion density increasing towards the center of the trap \cite{Dubin97}. This leads to a stronger Coulomb repulsion for the ions near the center of the chain. Correspondingly, the radial short-wavelength modes have amplitude which is larger at the center, and vanishes at the edges~\cite{morigi2004}. The structural instability is hence first visible at the center of the trap, such that at the critical value of the transverse frequency the central ions are displaced from the trap axis forming a zigzag chain, as illustrated in Fig.~\ref{zztrap}. At lower values of the transverse frequency the number of ions which are displaced from the center increases towards the edges of the chain, until all the chain is in a planar configuration~\cite{Schiffer93}. For a sufficiently long chain, one could associate with this behaviour a spatially-dependent critical frequency, determining the transition to the zigzag, such that it is largest at the center and smaller at the edges. Correspondingly, if a quench of the transverse frequency is applied, the transition point is crossed at different instant of times along the chain (from the center to the edges). The velocity, at which this front propagates, is the so-called front velocity~\cite{ikzm,ikzmions}. In this case, the ratio between the front velocity and the sound velocity determines nucleation of defects. In the following, we compute the number of kinks formed in an ion crystal in a linear Paul trap in such scenario. The mechanism resembles the formation of solitons in a cigar-shaped Bose-Einstein condensate recently discussed by Zurek \cite{Zurek09bec}.

%%%%%%%%%%%%%%%%%%%%%%%%%%%%%%%%%
%%%%%%%%%%%%%%%%%%%%%%%%%%%%%%%%%%%

More precisely, we now consider the case, in which the ions are also trapped in the $x$-direction by a harmonic potential of frequency $\nu$, such that the aspect ratio $\nu/\nu_t$ is sufficiently small to allow for low-dimensional crystalline structures. We will now consider the case in which the number $N$ of particles is finite, but it is sufficiently large to allow for a local density approximation. In this limit and away from the chain edges the linear density $n(x)$ is well approximated by the function~\cite{Dubin97}
\begin{equation}
n(x)=\frac{3}{4}\frac{N}{L}\left(1-\frac{x^2}{L^2}\right),
\end{equation}
with $L$ the half-length of the chain and $x$ the distance from the center. Within this treatment, the interparticle spacing $a(x)$ is a slowly-varying function of the position, such that $a(x)=1/n(x)$. In the thermodynamic limit, in which $a(0)$ is fixed as the number of particles goes to infinity,  $N\to\infty$, one recovers the statistical mechanics and dynamical properties of the ion chain in a infinite-radius ring trap~\cite{morigi2004,fishman2008}. For $N$ finite, the transition from a linear to a zigzag chain can be estimated with the value $\nu_t^{(c)}\approx 3N\nu/(4\sqrt{\log N})$, which was found by taking only nearest-neighbours coupling and where the corrections scale with powers of $1/\log N$~\cite{morigi2004}. Apart from a factor of order  unity, it  corresponds to the relation
\begin{equation}
\label{eq:nutccenter}
\nu_t^{(c)2}=\frac{7}{2}\zeta(3)\frac{Q^2}{m a(0)^3}
\end{equation}
with $a(0)=1/n(0)$.

In order to determine a Ginzburg-Landau equation for this case, we first assume that the spatial variation is very slow, such that sufficiently far away from the edges, the length scale over which the interparticle distance changes is much larger than the coarse graining length  $\delta x\gg a(x)$ with which we study the dynamics:
\begin{equation}
\delta x=a(x)/\left|\frac{da}{dx}\right|\gg a(x)
%a(x)/\left|\frac{da}{dx}\right|\gg\delta x
\end{equation}
 In this limit, we can make a slowly-varying ansatz for the short-wavelength eigenmodes of the linear chain, such that for a given eigenmode we extend the treatment for the homogeneous case starting from Eq.~(\ref{eq:normal}) and write $\sigma_n=\alpha_n{\rm e}^{{\rm i}kna}$, with $\alpha_n$ slowly-varying amplitude~\cite{morigi2004}. Within a local-density approximation, the ions contained in a region of size $\delta x$ and centered at position $x$, become unstable at the position-dependent critical transverse frequency given by
\begin{equation}
\nu_t^{(c)}(x)^2=\frac{7}{2}\zeta(3)\frac{Q^2}{m a(x)^3},
\end{equation}
which has the same form as the one for the homogeneous case but with the appropriate interparticle distance $a(x)$.
Moreover, extending to the non-homogeneous case the Lagrangian in Eq.~(\ref{Eq:GL}), we find the Lagrangian $L'=\int {\rm d}x{\mathcal L}'(x)$, with the Lagrangian density
\begin{eqnarray}
\label{eq:GLinhomogeneous}
{\mathcal L}'(x)&=&\frac{1}{2}\rho(x)\Bigl[(\partial_t\psi(x))^2-h(x)^2\left(\partial_x\psi(x)\right)^2 \nonumber\\
&-&\delta(x)\psi(x)^2 -{\mathcal A}(x)\psi(x)^4\Bigr].
\end{eqnarray}
In Eq.~(\ref{eq:GLinhomogeneous}), $\rho(x)=mn(x)$ is the linear mass density, and the spatial dependence of the coefficients $\delta(x)$, $h(x)$, and $\mathcal A (x)$ is found by using the position-dependent interparticle distance $a(x)$ in the corresponding formulas for the homogeneous case. When deriving the Lagrangian density of Eq.~(\ref{eq:GLinhomogeneous}) we have neglected the coupling between axial and transverse degrees of freedom, assuming very small fluctuations about the critical point.

In order to derive the Euler-Lagrange equation for the field, we assume that the linear density, and hence the interparticle distance $a(x)$, does not depend on the value of the transverse trap frequency and thus remains constant when quenching $\nu_t$ through the critical point. Terms proportional to the spatial gradient ${\rm d}a/{\rm d}x$ are also neglected, assuming that sufficiently far away from the edges the inequality $|\psi^{\prime\prime}(x)|\gg |\psi^{\prime}(x)a^{\prime}(x)/a(x)|$ holds (which is consistent with the approximation made for deriving the Lagrange density). Within this limit, the equation of motion for the field $\psi=\psi(x,t)$ reads
\begin{equation}
\partial_t^2\psi-h(x)^2\partial_x^2\psi+\eta \partial_t\psi+\delta(x,t)\psi+2{\mathcal A}(x)\psi^3=\varepsilon(t),
\end{equation}
where now
\begin{eqnarray}
\delta(x,t)&=&\nu_t(t)^2-\nu_t^{c}(x)^2\nonumber\\
&=&\nu_t^{(c)}(0)^2-\nu_t^{(c)}(x)^2-\delta_0\frac{t}{\tau_Q},
\end{eqnarray}
and we considered the time-dependence given in Eq.~(\ref{Eq:quench}).

We now consider defect formation when the value of the transverse frequency is quenched through the critical point. This will happen in the central region of the chain first, giving rise to a propagating front along the axis, whose coordinates $(x_F,t_F)$ satisfy $\delta(x_F,t_F)=0$.
The front velocity $v_F$, at which the instability propagates, can be found by taking the ratio between the characteristic length of the control parameter, $\left(\partial_x\delta(x,t)/\delta(x,t)\right)^{-1}$, over the characteristic time scale at which it changes, $\left(\partial_t\delta(x,t)/\delta(x,t)\right)^{-1}$. It takes the form
giving
\beqa
v_F\sim \frac{\partial_t\delta(x,t)}{\partial_x\delta(x,t)}.
\eeqa
An explicit dependence on the physical parameter can be found using the spatially-dependent critical frequency in the expression
\beqa
\nu_c^2(x)=\nu_c^2(0)[1-X^2]^3,
\eeqa
with $X=x/L$. For the front velocity one obtains
\beqa
v_F\sim\frac{\delta_0}{\tau_Q}\left|\frac{d\nu_c^2(x)}{dx}\right|_{x_F}^{-1}=\frac{L\delta_0}{6\nu_t^{(c)}(0)^2\tau_Q}\frac{1}{|X|}(1-X^2)^{-2}.
 \eeqa
Whenever the transition is homogeneous, $v_F$ becomes infinite and the standard KZM applies, allowing for the nucleation of defects in the whole system.
Otherwise, as we show in the following, nucleation of kinks will only take place in an restricted fraction of the chain, with length $2\hat{X_*}L$,  where the front velocity $v_F$ is larger than the characteristic velocity $\hat{v}_x$ with which a perturbation propagates along the chain at the freeze-out time. We will find that the scaling of kinks density with the quenching rate will be in this case different from the scaling found in the homogeneous case.

In order to determine the density of defects, we now evaluate the velocity $\hat{v}_x$ with which a perturbation propagates along the chain at the freeze-out time.
An upper-bound for $\hat{v}_x$ can be found by considering the ratio  of the correlation length $\hat{\xi}_x$
and the relaxation time $\hat{\tau}_x$ at $\hat{t}$,
\beqa
\hat{v}_x\sim\frac{\hat{\xi}_x}{\hat{\tau}_x}.
\eeqa
where the $x$-subindex underlines the spatial dependence due to the inhomogeneous nature of the system. This also corresponds to the speed of sound at the freeze-out point at an energy which is one over the relaxation time.
 In order to compute it, first note that the relative frequency can be written with reference to $t_F$ as $\delta(x,t)=-\delta_0(t-t_F)/\tau_Q$, where the spatial dependence is encoded in $t_F=\tau[\nu_t^{(c)}(0)^2-\nu_t^{(c)}(x)^2]/\delta_0$. One can find the instant $\hat{\mathrm{t}}$, relative to $t_F$,  at which the dynamics stops being adiabatic by equating the time scale $\delta/\dot{\delta}$ to the relaxation time $\tau_x=\eta/\delta(x,t)$.
In the overdamped regime,  $\hat{\mathrm{t}}=(\eta\tau_Q/\delta_0)^{1/2}$, which sets the freezed-out correlation length $\hat{\xi}_x=a\om_0/\sqrt{|\delta(x,\hat{\mathrm{t}})|}=a\om_0(\eta\delta_0/\tau_Q)^{-1/4}$.
Hence, the characteristic velocity of a perturbation becomes $\hat{v}_x=\hat{\xi}_x/\hat{\tau}_x=a\om_0(\delta_0/\eta^3\tau_Q)^{1/2}$.

The key insight in the IKZM is that nucleation of defects is only expected
whenever the front velocity is larger than $\hat{v}_x$, namely,
\beqa
v_F>\hat{v}_x,
\eeqa
so that spatially separated regions of the chain are causally disconnected. The violation of this inequality opens the possibility of driving a truly adiabatic transition. The reason is that the choice of the ground state in the broken symmetry phase is not independent in different regions of the system whenever $v_F$ is small enough with respect to $\hat{v}_x$. As a result the inhomogeneous nature of the transition allows for an adiabatic crossing, where a zigzag chain results after the quench with no defects. This holds even in thermodynamic limit and dramatically differs from the HKZM where defects would always be expected for infinite systems.

In the overdamped regime, nucleation of kinks is still possible whenever
\beqa
\frac{v_F}{\hat{v}_x}=\mathcal{A}_o\frac{1}{|X|}(1-X^2)^{-2}>1,
\eeqa
 with
%$X=x/L$ and
\beqa
\mathcal{A}_o=\frac{L}{6\nu_t^{(c)}(0)^2 a\om_0}\left(\frac{\eta\delta_0}{\tau_Q}\right)^{\frac{3}{4}}.
\eeqa
This will generally be fulfilled in a limited region of the system, $2\hat{X_*}L$, where the homogeneous KZM applies. One can estimate the effective size of this region where defect nucleation is possible by setting $v_F/\hat{v}_x=1$, and assuming $\hat X_*\ll 1$. Then, it follows that $\hat{X_*}_o\simeq\mathcal{A}_o$ which leads to the scaling law of the density of kinks
\beqa
\label{okzm}
d_o\sim \frac{2\hat{X_*}_o}{\hat{\xi}_o}=\frac{L }{3\nu_t^{(c)}(0)^2a^2\om_0^2} \frac{\eta\delta_0 }{\tau_Q}.
\eeqa
Though the absolute density of defects is reduced with respect to the homogeneous case (since $\hat{X_*}< 1$), the dependence on the quenching rate is enhanced. Nonetheless, such IKZM scaling breaks down for fast quenches due to the finite size of the chain, and for very slow quenches due to relevant defects losses as will be discussed below in more detail.

An analogous description applies to the underdamped case, where the relaxation time is independent of the dissipation and diverges as $\tau_x=1/\sqrt{|\delta(x,t)|}$, leading to the freeze-out time $\hat{\mathrm{t}}=(\tau_Q/\delta_0)^{1/3}$. In this time scale, the correlation length freezes, with respect to the quench time scale $\tau_Q$, at a value $\hat{\xi}_x=a\om_0(\tau_Q/\delta_0)^{1/3}$
leading to a uniform sound velocity $\hat{v}_x=\hat{\xi}_x/\hat{\tau}_x=a\om_0$. Again, the transition remains adiabatic as long as
\beqa
\frac{v_F}{\hat{v}_x}=\mathcal{A}_u\frac{1}{|X|}(1-X^2)^{-2}>1,
\eeqa
where we introduced the parameter
\beqa
\mathcal{A}_u=\frac{L}{6\nu_t^{(c)}(0)^2 a\om_0}\frac{\delta_0 }{\tau_Q}.
\eeqa
Using the same argument employed in the overdamped regime, kinks are found in a region of size $\hat{X_*}_u\simeq\mathcal{A}_u$, so that
\beqa
\label{ukzm}
d_u\sim\frac{2\hat{X_*}_u}{\hat{\xi}_u}
=\frac{L}{3\nu_t^{(c)}(0)^2a^2\om_0^2}\left(\frac{\delta_0 }{\tau_Q}\right)^{4/3}.
\eeqa

\begin{figure}[t]
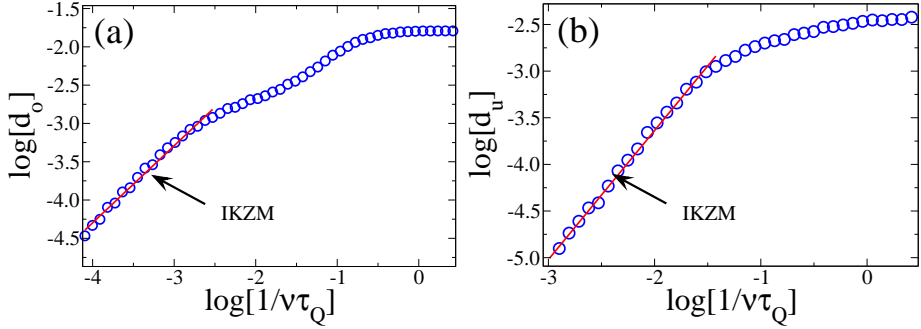

\begin{center}
\includegraphics[width=6cm,angle=0]{longkzm_fig4a.eps}
\includegraphics[width=6cm,angle=0]{longkzm_fig4b.eps}
\caption{\label{IKZM}
Density of defects  for a harmonically trapped ion chain as a function of the inverse of the sweeping rate
(a) in the overdamped regime ($\eta=100\nu$), where the slope in the fit is $1.006$ with regression coefficient $0.995$.
(b) in the underdamped regime ($\eta=10\nu$), where the slope in the fit is $1.384$ with regression coefficient $0.997$.
The defects are only considered in the central $N_\mathcal{C}=30$ ions, in order to minimize defect losses
($N=50$, $2000$ realizations). The parameters are  $\nu_t^{(c)}(0)\simeq 18\nu$,
$\delta_0=36\nu^2$, $\epsilon=0.05 l_0\nu^{3/2}$, with $l_0^3=Q^2/m\nu^2$ and $\nu$ being the axial frequency.
}
\end{center}
\end{figure}

To test the IKZM, we study numerically in Figure \ref{IKZM} the scaling of the density of kinks as a function of the quenching rate both in the overdamped and underdamped regimes. The results show a good agreement with the predictions in Eqs. (\ref{okzm}) and (\ref{ukzm}). Nonetheless, there is a saturation of the density of kinks at high quenching rates due to the interactions between the kinks.
%
%
 %%%%%%%%%%%%%%%%%%%%%%%%%%%%%%%%%%%%%%%%%%%%%%%%%%%%%%%%%%%%%%%%%%%%%%%%%%%%%%
%%%%%%%%%%%%%%%%%%%%%%%%%%%%%%%%%%%%%%%%%%%%%%%%%%%%%%%%%%%%%%%%%%%%%%%%%%%%%%

We  further note that the ratio between the number of defects $\mathcal{N}$ that nucleate according to the IKZM  and HKZM obeys the relation
\beqa
\frac{\mathcal{N}_{IKZM}}{\mathcal{N}_{HKZM}}=2\hat{X_*},
\eeqa
with a well define scaling with respect to the quenching rate, and which applies both in the underdamped and overdamped regimes.
In addition, the HKZM is known to overestimate the number of defects by a numerical factor $f$ of order unity, $f\sim 4.5$ in our case.

\subsection{Discussion}
In this section we discuss different mechanisms responsible for deviations from the IKZM scaling.
Two new effects come into play with respect to the ring configuration: a) coupling between axial and transverse modes b) enhanced defect transport.
We shall dwell on the implications of these two effects in KZM scaling.

{\it Axial-transverse mode coupling} - The (inhomogeneous) trapping potential allows the ions to shift in the axial direction as the structural phase transition takes place. As a result, during the course of the transition the axial density of ions increases near the center of the trap,
an effect which hinders the study of the IKZM, as expected from the derivation of the GLE in the thermodynamic limit and further suggested by numerical simulations.

{\it Dynamical losses of defects} - In order to minimize the axial-transverse mode coupling it is desirable to drive the transition just in the center of the chain, so that the ions at the edges of the trap remain in the linear configuration. For the ground state, the amplitude of the transverse displacement of the ions increases monotonically from each of the edges of the chain towards the center of the trap. Hence, the effective Peierls-Nabarro potential \cite{PN,FKM} seen by a kink decays as one approaches the ends of the chain. As a consequence there is transport of defects, which provides a mechanism for their losses near the edges of the trap.  Note that defect transport remains even if the longitudinal degrees of freedom of the ions are frozen on a lattice: the transverse motion suffices for its dynamics.
A way of minimizing these defect losses is by making the inter-ion spacing homogeneous, which in turn makes
the Peierls-Nabarro potential periodic along the chain.
Different potentials have been proposed to achieve homogeneous inter-ion spacing \cite{Harald}. For an ion chain in a linear Paul trap, the optimal axial trapping potential can be found by fitting the local Coulomb potential in an homogeneous chain, $U(\{x_n\})=\sum_{n\neq n'}\frac{1}{|x_n-x_{n'}|}=\frac{1}{a}\sum_{n\neq n'}\frac{1}{|n-n'|}$ where $a$ is the desired value of the inter-ion spacing. Under such axial confinement, the transition becomes then more homogeneous. Nonetheless, there is a local correction to the transverse critical frequency, $\nu_{r,eff}^2(n)=\nu_r(t)^2-\sum_{n\neq n'}\frac{e^2/m}{|x_n-x_{n'}|^3}$ which varies as a function of the position along the chain. Hence even when the ions are homogeneously spaced, the transition remains inhomogeneous.
The underlying defect dynamics is also driven by the fact that pairs of defects with the same topological charge repel each other and attract otherwise. Scattering between kinks and anti-kinks (of positive and negative topological charge) can occur leading to their annihilation. Nonetheless defects stop seeing each other when they are separated by few ions ($\sim 5$), which motivates the study of thes scaling with moderate densities. 
Under the conditions in Fig. (\ref{IKZM}) defects are stable on a scale $\sim 10\nu^{-1}$ which suffices for its imaging in the laboratory. 
The mechanisms for defect losses are particularly relevant in the underdamped regime, where higher scaling coefficients than those predicted by the KZM are observed
and the fluctuations in the number of kinks increase. Generally, whenever defects losses are relevant the connection of the scaling of the density of defects with KZM can be questioned,
since these process are non-universal and disregarded in the mechanism. As a result, the overdamped regime is single out as optimal for studies of the IKZM.
 
We close by noticing that the minimal length of the chain required to verify the IKZM scaling is constrained by two
finite-size effects:
a) The existence of the adiabatic dynamics whenever the correlation length at the freeze out time, $\hat{\xi}=\xi(\hat{t})$, equals the length of the part of the  chain where
defects are counted, and more fundamentally, where defects can nucleate $2\hat{X}_*$.
Defects nucleate as long as $2\hat{X}_*L>\hat{\xi}�$.
b) The breakdown of the IKZM scaling at fast rates due to a saturation of the average density of defects. If one wishes to check the scaling by varying the quenching time between $\tau_i$ and $\tau_f$ (ideally ranging over few  orders of magnitude), it should be possible to achieve the corresponding densities of defects ($d_i$ and $d_f$) related as $d_i=(\tau_f/\tau_i)^\alpha d_f$, where $\alpha$ is the IKZM scaling  ($1$ in the overdamped regime, $4/3$ in the underdamped regime), without jumping into the adiabatic dynamics. Long enough chains were created in various setups. In \cite{MPQ} a chain of up to $5 \times 10^4$ ions was created with an axial trapping frequency of $1$ MHz. Ring configuration were created dynamically in \cite{Schaetz}. In the same setup static structures of up to $50$ ions were created at the temperature of $1$ mK.

\section{Conclusions}
In this work we analyzed the formation of defects in a one dimensional Coulomb crystal during a frequency quench from the linear chain to the zigzag structure. We studied the cases of ions crystals in a linear Paul trap and in a ring-shaped trap  and predicted the defects production rate. Our study shows that a Coulomb crystal is a particularly neat and controllable system where the homogeneous and inhomogeneous KZM can be tested.
Despite the experimental work in \cite{kzmexp} addressing the homogeneous KZM (HKZM), the inhomogeneous KZM (IKZM) \cite{ikzm,ikzmions} lacks to date experimental verification,
and the ion chains in a Paul trap are put forward here as an ideal system for such goal.

{\it Acknowledgements.} We thank T. Calarco, S. Fishman,  H. Rieger, H. Landa, S.Marcovitch and B. Reznik for fruitful discussions and R. Rivers and J. Dziarmaga for useful comments.
We further ackowledge support by the European Commission (AQUTE, SCALA and QAP,
STREPs HIP and PICC), the EPSRC, ESF EUROQUAM CMMC, the Generalitat de Catalunya Grant No. 2005SGR-00343 and the Spanish Ministerio de Educaci\'on y Ciencia (FIS2007-66944; FIS2008-01236;
Juan de la Cierva; Ramon-y-Cajal, Consolider Ingenio 2010 "QOIT").
G.M. and M.P. acknowledge the support of a Heisenberg Professorship and an Alexander-von-Humboldt Professorship, respectively.

\appendix 
\section{Relaxation time}

In order to compute the relaxation time, we consider the following autocorrelation function:
\begin{eqnarray}
\label{gtau}
g_t(\tau)=\langle y_n(t+\tau)y_n(t) \rangle = \frac{1}{N}\sum_{kk'} e^{i(k+k') na }\langle \Psi_k(t+\tau)\Psi_{k'}(t) \rangle 
\end{eqnarray}
where for simplicity we just study the dynamics of the ions in one direction $y$ and we inverted Eq. (\ref{eq:normal}) for the expansion of the particles coordinates in terms of normal modes.
In the overdamped regime: $\eta\gg \sqrt{|\beta(k)|}$ we can neglect the second order derivative in Eq. (\ref{eq:langevink}) whose solution becomes
\begin{eqnarray}
\Psi_k(t)= \Psi_k e^{-\frac{\beta(k)}{\eta} t} +\frac{1}{\eta}\int_0^t  e^{-\frac{\beta(k)}{\eta} (t-s)} \varepsilon_k(s) ds
\end{eqnarray}
Substituting this solution in the definition of $g_t(\tau)$ Eq.(\ref{gtau}) we get:
\begin{equation}
g_t(\tau)=\frac{1}{N} \sum_k e^{-\frac{\beta(k)}{\eta}(2t+\tau)}\left[\langle\Psi_k\Psi_{-k} \rangle -\frac{2\kappa_B T}{m\beta(k)}\left(e^{-\frac{2\beta(k)}{\eta}t} -1\right) \right]
\end{equation}
Now for fixed time $t$, all the components with different momentum in  the autocorrelation function $g_t(\tau)$ decay exponentially with the time separation $\tau$. The largest time scale, corresponding to the minimum frequency $\delta=\min_k \beta(k)$, defines the relaxation time:
\begin{equation}
\tau_o= \frac{\eta}{\delta}
\end{equation}
where the subscript $o$ stresses that this result has been obtained in the overdamped regime. Note also that defects are localised excitations and a correction depending on $h$ can be obtained from the two-point correlation function.


\section*{References}
\begin{thebibliography}{10}
\bibitem{DubinRMP}
Dubin D H E and  O'Neil T M 1999 {\it Rev. Mod. Phys.} {\bf 71} 87

\bibitem{MPQ}
%Multiple shells structures of laser-cooled 24Mg+ ions in a quadrupole storage
%ring
Birkl G, Kassner S and Walther H (1992) {\it Nature} {\bf 357} 310;
%Observation of ordered structures of laser-cooled ions in a quadrupole
%storage ring,
Waki I, Kassner S, Birkl G and Walther H (1992) {\it Phys. Rev. Lett.} {\bf 68} 2007.

\bibitem{Raizen}
%Ionic crystals in a linear Paul trap,
Raizen M G, Gilligan J M, Bergquist J C, Itano W M and Wineland D J (1992) {\it Phys.\ Rev.} A {\bf 45} 6493


\bibitem{Werth}
%Crystalline ion structures in a Paul trap
Block M, Drakoudis A, Leuthner H, Seibert P and Werth G (2000), {\it J. Phys.} B{\it : At. Mol. Opt. Phys.} \textbf{33} L375

\bibitem{Drewsen}
%Observation of strutcural transition for Coulomb crystals in a linear Paul
%trap
Kjargaard N and Drewsen M (2003) {\it Phys. Rev. Lett.} {\bf 91} 095002;
%Observation of Three-Dimensional Long-Range Order in Small Ion Coulomb Crystals in an rf Trap
Mortensen A, Nielsen E, Matthey T and Drewsen M (2006) {\it Phys. Rev. Lett.} \textbf{96} 103001.

\bibitem{Bollinger}
%Bragg Diffraction from Crystallized Ion Plasmas
Itano W M, Bollinger J J, Tan J N, Jelenkovic B, Huang X P and Wineland D J (1998) {\it Science} {\bf 279} 686

\bibitem{Schiffer93} Schiffer J P (1993) {\it Phys. Rev. Lett.} \textbf{70} 818

\bibitem{Piacente2005}
Piacente G, Schweigert I V, Betouras J J and Peeters F M (2004) {\it Phys. Rev.} B {\bf 69} 17

\bibitem{fishman2008}
%Structural phase transitions in low dimensional ion crystals
Fishman S, De Chiara G, Calarco T and Morigi G (2008) {\it Phys. Rev.} B {\bf 77} 064111
\bibitem{ramsey} De Chiara G, Calarco T, Fishman S,  Morigi G (2008) {\it Phys. Rev.} A {\bf 78} 043414 
\bibitem{RTSP} Retzker A, Thompson R C, Segal D M and Plenio M B (2008) {\it Phys. Rev. Lett.} \textbf{101} 260504

\bibitem{Schuetzhold} Uhlmann M, Sch\"utzhold R and Fischer U R (2010), {\it Phys. Rev.} D {\bf 81} 025017; arXiv:1005.2649

\bibitem{landa}  Landa H, Marcovitch S, Retzker A, Plenio M B and Reznik B (2010) {\it Phys. Rev. Lett.} {\bf 104} 043004

\bibitem{Landau} Landau L D, Lifshitz E M and Pitaevskii L P (1980) {\it Statistical Physics, 3rd ed.}
(Oxford: Reed Educational and Professional Publishing Ltd.)

\bibitem{HHreview} Hohenberg P C and Halperin B I (1977) {\it Rev. Mod. Phys.} {\bf 49} 435

\bibitem{Kibble} Kibble T W B (1976) {\it J. Phys. A: Math. Gen.} {\bf 9} 1387; (1980) {\it Phys. Rep.} {\bf 67} 183

\bibitem{Zurek}
Zurek W H (1985) {\it Nature (London)} {\bf 317} 505; (1993) {\it Acta Phys. Pol.} B {\bf 24} 1301

\bibitem{kzmnum} Laguna P and Zurek W H (1997) {\it Phys. Rev. Lett.} {\bf 78} 2519;
%Phys. Rev. D 58, 5021 (1998);
Yates A and Zurek W H (1998) {\it Phys. Rev. Lett.} {\bf 80}, 5477;
Stephens G J, Calzetta E A, Hu B L and Ramsey S A (1999) {\it Phys. Rev. D} {\bf 59} 045009;
Antunes N D, Bettencourt L M A and Zurek W H (1999) {\it Phys. Rev. Lett.} {\bf 82} 2824;
%J. Dziarmaga, P. Laguna and W. H. Zurek, {\it ibid}. {\bf 82}, 4749 (1999);
Hindmarsh M B and Rajantie A (2000) {\it ibid}. {\bf 85} 4660;
Stephens G J, Bettencourt L M A and Zurek W H (2002) {\it ibid}. {\bf 88}, 137004

\bibitem{kzmexp}
Chuang I, Durrer R, Turok N and Yurke B (1991) {\it Science} {\bf 251} 1336;
Bowick M J, Chandar L, Schiff E A and Srivastava A M (1994) {\it ibid}. {\bf 263} 943;
Ruutu V M H, Eltsov V B, Gill A J, Kibble T W B, Krusius M, Makhlin Yu G, Pla{\c c}ais B, Volovik G E and Xu Wen (1996) {\it Nature} {\bf 382}, 334;
B\"auerle C, Bunkov Y, Fisher S N, Godfrin H and Pickett G R (1996) {\it ibid}. {\bf 382}, 332;
Carmi R, Polturak E and Koren G (2000) {\it Phys. Rev. Lett.} {\bf 84} 4966;
Maniv A, Polturak E and Koren G (2003), {\it ibid}. {\bf 91}, 197001 (2003);
Monaco R, Mygind J and Rivers R J (2002) {\it Phys. Rev. Lett.} {\bf 89} 080603; (2003) {\it Phys. Rev.} B {\bf 67} 104506;
(2006) {\it Phys. Rev. Lett.} {\bf 96} 180604; (2008) {\it Phys. Rev.} B {\bf 77} 054509;
Ducci S, Ramazza P L, Gonz\'alez-Vi\~{n}as W and Arecchi F T (1999) {\it Phys. Rev. Lett.} {\bf 83} 5210;
 Casado S, Gonz\'alez-Vi\~{n}as W, Mancini H and Boccaletti S (2001) {\it Phys. Rev.} E {\bf 63}, 057301;
 (2006) {\it ibid}. {\bf 74} 047101; (2007) {\it Eur. J. Phys.} {\bf 146} 87


\bibitem{kzmqpt}
Damski B (2005) {\it Phys. Rev. Lett.} {\bf 95} 035701;
Zurek W H, Dorner U and Zoller P (2005) {\it Phys. Rev. Lett.} {\bf 95} 105701;
Dziarmaga J (2005) {\it Phys. Rev. Lett.} {\bf 95} 245701;
Polkovnikov A (2005) {\it Phys. Rev.} B {\bf 72} 161201(R)



\bibitem{Zurek09bec} Zurek  W H (2009) {\it Phys. Rev. Lett.} {\bf 10} 105702
\bibitem{ikzm}  Kibble T W B and Volovik G E (1997) {\it Pis'ma v ZhERF} {\bf 65} 96;
Dziarmaga J, Laguna P, Zurek W H (1999) {\it Phys. Rev. Lett.} {\bf 82} 4749 (1999);
Dziarmaga J and Rams M M, (2010) {\it New J. Phys.} {\bf 12}, 055007.
\bibitem{ikzmions} del Campo A, De Chiara G, Morigi G, Plenio M B, and Retzker A (2010) {\it Phys. Rev. Lett.}  {\bf 105}, 075701.


\bibitem{morigi2004}

%Eigenmodes and thermodynamics of ion chain in a harmonic potential
Morigi G and Fishman S (2004) {\it Phys. Rev. Lett.} {\bf 93} 170602;
%Dynamics of an ion chain in a harmonic potential
(2004) {\it Phys. Rev.} E {\bf 70} 066141

\bibitem{morigi2001}
%Doppler cooling of a Coulomb crystal
Morigi G and Eschner J (2001) {\it Phys. Rev.} A {\bf 64} 063407


\bibitem{Footnote:2}
Equation~(\ref{eq:langevink}) is in general valid even if few ions of the crystal are driven by the laser. This is correct provided that the system is far enough from the transition point. In fact, the cooling rate scales with the recoil frequency of the ion, which is smaller than the frequency $\omega_0$ characterizing the Coulomb coupling between the ions in the axial direction and thus the time in which perturbation can propagate from one point of the chain to any other (In this case the damping rate $\eta_{\sigma,k}$ depends also on the corresponding mode spatial mode structure at the points where the ions are illuminated~\protect{\cite{MorigiEPJD01}}). This will no more be true  when approaching the transition, such that the typical time scales of the radial normal modes may become very small. Hence, when cooling only few ions of the chain in this regime, the mechanical effects of light may produce local deformations of the chain structure.

\bibitem{MorigiEPJD01}
Morigi G and Walther H (2001) {\it Eur. Phys. J.} D {\bf 13} 261

\bibitem{laguna1998}
%Critical dynamics of symmetry breaking: Quenches, dissipation, and cosmology
Laguna P and Zurek W H (1998) {\it Phys. Rev.} D {\bf 58} 085021

\bibitem{FKM} Barun O M and Kivshar Y S (2004) {\it the Frenkel-Kontorova model} (Heidelberg, Springer);
Braun O M, Chubykalo O A, Kivshar Y S, V\'azquez L (1993) {\it Phys. Rev.} B {\bf 48} 3734

\bibitem{Dubin97} Dubin D H E (1997) {\it Phys. Rev.} D {\bf 55} 4017

\bibitem{PN}%the size of a dislocation
 Peierls R (1940) {\it Proc. phys. Soc.} {\bf 52} 34;
%dislocations in a simple cubic lattice
Nabarro F R N (1947) {\it Proc. phys. Soc.} {\bf 59} 256

\bibitem{Harald} Wunderlich H, Wunderlich C, Singer K, Schmidt-Kaler F (2009) {\it Phys. Rev.} A {\bf 79} 052324;
Lin G D, Zhu S L, Islam R, Kim K, Chang M S, Korenblit S, Monroe C, Duan L-M (2009) {\it Europhys. Lett.} {\bf 86} 60004

\bibitem{Schaetz} Sch\"atz T, Schramm U, and Habs D (2001) {\it Nature (London)} {\bf 412} 717
\end{thebibliography}
\end{document}